\documentclass[preprint,12pt]{elsarticle}

\usepackage{amssymb}
\usepackage{amsmath}

\usepackage{amsfonts}
\usepackage{bm}
\usepackage{booktabs}
\usepackage{geometry}
\usepackage{hyperref}

\hypersetup{
	colorlinks=true,
	linkcolor=blue,
	citecolor=blue,
	filecolor=magenta,      
	urlcolor=cyan,
}

\journal{Physics Letters B}

\begin{document}

\begin{frontmatter}



\title{Exact 
model of a gravitational wave and electromagnetic radiation in DHOST gravity theory: wave solutions and testable effects}


\author[O1,O2]{Konstantin E. Osetrin} 

            
\affiliation[O1]{organization={Center for Theoretical Physics, 
Faculty of Physics and Mathematics, Tomsk State Pedagogical University},
addressline={Kievskaya, 60},
city={Tomsk},
postcode={634041},
country={Russia}}

\affiliation[O2]{organization={International Laboratory for Theoretical Cosmology, 
Tomsk State University of Control Systems and Radioelectronics},
addressline={Lenina pr. 40},
city={Tomsk},
postcode={634050},
country={Russia}}

\begin{abstract}
Within an analytical non-perturbative framework based on a single exact model, the coupled dynamics of gravitational waves and 
high-frequency electromagnetic fields is investigated in General Relativity and DHOST gravity. 
Using an exact gravitational-wave spacetime in an isotropic (null) wave coordinate system, Maxwell's equations are integrated analytically. 
This analysis provides an exact non-perturbative confirmation from the full field equations that a gravitational wave acts as a waveguide, 
forming a propagation channel for electromagnetic radiation co-directional with the gravitational wave and generating an effective pure-radiation source. 
Applying this exact electrodynamic solution to the nonlinear field equations of DHOST class~Ia gravity reveals 
that the algebraic structure of the DHOST constraint equations enforces a modification of the electromagnetic-wave parameters. 
Based on this mechanism, an observationally testable effect of scalar frequency drift is derived, 
providing a direct way to discriminate between General Relativity and modified gravity theories 
even when their metric tensors share identical functional forms.
\end{abstract}



\begin{highlights}
	\item An end-to-end exact method is developed to test modified gravity theories.
	\item Exact coupled wave solutions are found from Maxwell theory to GR and DHOST gravity.
	\item Exact field equations confirm gravitational waves form a co-directional light channel.
	\item Exact scalar frequency drift formulas are derived in DHOST gravity.
	\item New spectroscopic and polarization tests distinguish kinematically identical metrics.
\end{highlights}


\end{frontmatter}



%
%
%
%

\section{Introduction}
\label{sec:intro}

The construction of cosmological models alternative to the standard $\Lambda$CDM paradigm currently faces stringent constraints imposed by multi-messenger astronomy data. Gravitational-wave surveys (including the current observations of the LIGO-Virgo-KAGRA network) have imposed the most stringent bound on the speed of gravitational tensor perturbations $c_g$, establishing its equality with the speed of light at the level of the fifteenth decimal place \cite{LIGOScientific:2026uyd}. This has led to the exclusion of many modified gravity theories (MGT).

Among scalar-tensor gravity theories (STT), the most viable family that remains consists of the degenerate higher-order theories  -  the DHOST (Degenerate Higher-Order Scalar-Tensor) class~Ia theories, which, owing to the degeneracy conditions, are free from instabilities and automatically ensure the equality $c_g = c$. Moreover, data from spectroscopic surveys of the large-scale structure of the Universe obtained with the DESI instrument indicate the possibility of a dynamical evolution of the dark-energy equation of state, which provides additional motivation for studying models with evolving nonlinear scalar backgrounds $\phi(x^\alpha)$ (see \cite{DESI:2025fxa,Zhu:2026koz,Odintsov:2025jfq}).

The most general class of scalar-tensor theories (STT) leading to equations of motion with no higher than second-order differential equations was long considered to be ``Horndeski gravity'' \cite{Horndeski1974,Kobayashi:2019hrl}, later generalized to the subclass of ``Galileon models''~\cite{Deffayet:2009wt}. The absence of higher derivatives in the field equations was a strict mathematical requirement (the Ostrogradsky criterion) for preventing instabilities in the form of negative-energy modes.

It was shown in Ref.~\cite{Langlois:2015cwa} that the Horndeski criterion is sufficient but not necessary. The authors demonstrated that Lagrangians containing second derivatives of the scalar field, $\phi_{\alpha \beta }=\nabla_\alpha \nabla_\beta  \phi$, can lead to physically consistent equations free of instabilities, provided that the matrix of their kinetic terms is degenerate. This fundamental class of theories became known as DHOST (Degenerate Higher-Order Scalar-Tensor) theories. The degeneracy of the kinetic matrix guarantees the existence of a constraint that eliminates the parasitic Ostrogradsky degree of freedom, leaving the theory with exactly three physical degrees of freedom (two tensor modes and one scalar mode), as in Horndeski theory.

The classification of DHOST Lagrangians quadratic in the higher derivatives was presented in Refs.~\cite{Crisostomi:2016czh,Crisostomi:2017ugk} and  \cite{Langlois:2015cwa,Langlois:2018dxi}, where the space of degenerate theories was divided into several isolated classes. 

The choice of the specific DHOST class~Ia in the present work is dictated by stringent cosmological constraints formulated after the coincident detection of gravitational and electromagnetic waves from the binary neutron-star merger GW170817~\cite{LIGOScientific:2017vwq}. 
As shown in Refs.~\cite{Crisostomi:2017lbg,Ezquiaga:2017ekz}, the requirement of the equality of the speeds of gravitational and electromagnetic waves has excluded most of the parameter space of STT. 
Class~Ia turned out to be a nontrivial subclass of higher-order DHOST theories that preserves the condition $c_g = c$ on cosmological scales throughout the entire evolution of the Universe, and in addition incorporates an efficient mechanism of nonlinear Vainshtein screening \cite{Vainshtein:1972sx,Babichev:2013usa}, which inside compact astrophysical systems (the Solar System, binary pulsars) suppresses the modified ``fifth force'', forcing the theory to mimic General Relativity to high accuracy in agreement with local tests.

The action and field equations for the quadratic DHOST~Ia Lagrangian were presented in Refs.~\cite{Langlois:2015cwa,BenAchour2016Dhost,Crisostomi:2017ugk,Langlois_2021,Bellorin_2026}.

The traditional perturbative approach, which treats small gravitational-wave perturbations on the background of a homogeneous and isotropic Friedmann--Robertson--Walker universe, often masks the fundamental properties of MGT in strong-field regimes and at early stages of the evolution of the universe. 
An alternative is to turn to exact, non-perturbative solutions of the gravitational field equations. Of particular interest are exact wave models describing gravitational waves of arbitrary amplitude.
A distinguished class is that of the wave spacetimes of Petrov type~N, for which the curvature scalar vanishes identically, $R \equiv 0$. 
In General Relativity (GR) such metrics describe gravitational pp-waves in vacuum, with a vanishing Ricci tensor ($R_{\alpha\beta} = 0$).
These models include, in particular, exact models of gravitational waves in Bianchi type~IV, VI and VII universes
\cite{Osetrin2022EPJP856,Osetrin2022894,Osetrin325205JPA_2023,Obukhov:2024irh}. 

The consistent material source for these spacetimes is pure radiation, whose energy-momentum tensor (EMT) has the co-directed null structure $T_{\alpha\beta} = \rho\, l_\alpha l_\beta$.
However, the usual hydrodynamic approach to pure radiation as a null fluid is phenomenological and lacks wave characteristics such as frequency and phase. To overcome this problem, in the present work we make a transition to microscopic wave electrodynamics: we use a high-frequency electromagnetic wave as an analytical non-perturbative solution of Maxwell's equations directly on the curved background of the exact model of a gravitational wave of arbitrary amplitude. 

The aim of the study is to obtain and solve a self-consistent system of Maxwell's equations and DHOST theory in an isotropic wave coordinate system. We show that the nonlinear scalar sector of DHOST breaks the apparent equivalence of the metrics in the GR and DHOST theories, leading to a change in the frequency and phase characteristics of the propagating light, and on this basis we formulate astrophysical tests that can be verified by astronomical observations.

\section{Exact solutions of Maxwell's equations on the background of a gravitational pp-wave}
\label{sec:maxwell}

We choose an isotropic (null) coordinate system $x^\alpha  = \{u, v, x, y\}$, where $u$ and $v$ are null variables. The metric of a gravitational wave with signature $\{+, -, -, -\}$ is given by the expression:
\begin{equation}
	g^{\alpha \beta } = \begin{pmatrix} 0 & 1 & 0 & 0 \\ 1 & 0 & 0 & 0 \\ 0 & 0 & g^{22}(u) & g^{23}(u) \\ 0 & 0 & g^{23}(u) & g^{33}(u) \end{pmatrix},
	\qquad
	\alpha, \beta, \gamma=0,1,2,3,
\end{equation}
where the transverse anisotropic block $g^{AB}(u)$ ($A, B = 2, 3$) depends only on the null coordinate $u$, which plays the role of the wavefront of the gravitational wave. 

To describe high-frequency electromagnetic radiation we use a 4-potential of the form $A_\alpha  = \{A_u, A_v, A_x, A_y\}$. 
We fix a null transverse gauge:
\begin{equation}
	A_u = 0, \quad A_v = 0 \implies A_\alpha  = \bigl\{0, 0, A_2(u,x,y), A_3(u,x,y)\bigr\}.
\end{equation}
The covariant electromagnetic field tensor is defined as 
\begin{equation}
	F_{\alpha \beta } = \frac{\partial A_\beta }{\partial x^\alpha } - \frac{\partial A_\alpha }{\partial x^\beta }.
\end{equation}
We take the potential of a running wave in the real form, with the phase 
\begin{equation}
	\psi = \omega u + k_x x + k_y y,
\end{equation}
which gives the following expressions for the potentials:
\begin{equation}
	A_2 = a_2(u) \cos\psi, \quad A_3 = a_3(u) \cos\psi,
\end{equation}
where $\omega$ is the frequency and $k_x, k_y$ are the wave numbers in the transverse plane. 

Direct differentiation yields the components of the electromagnetic field tensor $F_{\alpha \beta }$:
\begin{align}
	F_{01} &\equiv F_{uv} = 0, \quad F_{1\beta } \equiv F_{v\beta } = 0, \\
	F_{02} &\equiv F_{ux} = -\omega a_2(u) \sin\psi + a'_2(u) \cos\psi, \\
	F_{03} &\equiv F_{uy} = -\omega a_3(u) \sin\psi + a'_3(u) \cos\psi, \\
	F_{23} &\equiv F_{xy} = (k_y a_2(u) - k_x a_3(u)) \sin\psi.
\end{align}
Here and below a prime denotes differentiation with respect to the coordinate $u$.
The matrix of covariant components of the electromagnetic field tensor takes the form:
\begin{equation}
	F_{\alpha \beta } = \begin{pmatrix} 0 & 0 & F_{02} & F_{03} \\ 0 & 0 & 0 & 0 \\ -F_{02} & 0 & 0 & F_{23} \\ -F_{03} & 0 & -F_{23} & 0 \end{pmatrix}.
\end{equation}

The second pair of Maxwell's equations in vacuum, 
\begin{equation}
	\frac{\partial}{\partial x^\beta }\left(\sqrt{-g}F^{\alpha \beta}\right) = 0,
\end{equation}
splits into four equations:
\begin{align}
	\alpha =u: \quad & 0 = 0, \\
	\alpha =v: \quad &  \omega \cos\psi \left(a_2 (k_x g^{22} + k_y g^{23}) + a_3 (k_x g^{23} + k_y g^{33})\right)  \nonumber \\
	&  + \sin\psi \left(g^{22} a'_2 + g^{23}a'_2 + g^{23}a'_3 + g^{33}a'_3\right)  = 0, \label{eq:max_v} \\
	\alpha =x: \quad & -k_y (k_y a_2(u) - k_x a_3(u))  \cos\psi = 0, \label{eq:max_x} \\
	\alpha =y: \quad & -k_x (-k_y a_2(u) + k_x a_3(u))  \cos\psi = 0. \label{eq:max_y}
\end{align}

Analysis of Eqs.~\eqref{eq:max_x} and \eqref{eq:max_y} shows that their strict fulfillment in vacuum requires the vanishing of the expression:
\begin{equation}
	k_y a_2(u) - k_x a_3(u) = 0 \implies F_{23} \equiv 0.
\end{equation}
For clarity, let us pass from the isotropic wave coordinate system used here to the physical time $t$ and the spatial axis $z$ of light propagation:
\begin{equation}
	u = \frac{t + z}{\sqrt{2}}, \quad v = \frac{t - z}{\sqrt{2}}.
\end{equation}
In such a chronogeometric basis $\{t, x, y, z\}$ the components of the magnetic field are computed as $\vec{B} = \text{rot}\,\vec{A}$. The longitudinal component of the magnetic field along the wave-propagation axis $z$ proves to be equal to:
\begin{equation}
	B_z = \frac{\partial A_y}{\partial x} - \frac{\partial A_x}{\partial y} = F_{23} = (k_y a_2 - k_x a_3) \sin\psi.
\end{equation}
Thus, the vacuum Maxwell equations \eqref{eq:max_x}--\eqref{eq:max_y} require the longitudinal magnetic field to vanish identically ($B_z \equiv 0$), which is equivalent to a strict potentiality condition for the wavefront. At the same time, standard electromagnetic wave fields are induced in the transverse sector:
\begin{align}
	E_x = \frac{1}{\sqrt{2}}(-\omega a_2 \sin\psi + a'_2 \cos\psi), \quad B_x = \frac{1}{\sqrt{2}}(\omega a_3 \sin\psi - a'_3 \cos\psi), \\
	E_y = \frac{1}{\sqrt{2}}(-\omega a_3 \sin\psi + a'_3 \cos\psi), \quad B_y = \frac{1}{\sqrt{2}}(-\omega a_2 \sin\psi + a'_2 \cos\psi).
\end{align}

The only nontrivial solution that turns Eq.~\eqref{eq:max_v} into an identity for any background $g^{AB}(u)$ is the case $k_x = 0$, $k_y = 0$, in which the wave vector of the electromagnetic radiation is strictly aligned with the covariantly constant Killing vector of the gravitational wave $L^\alpha  = (0,1,0,0)$, which defines its direction of propagation. 
Thus, in order not to violate the symmetry of the exact gravitational-wave model, the electromagnetic wave must propagate along the same direction as the gravitational wave (channeling of radiation).
In this regime the phase becomes strictly flat, 
\begin{equation}
	\psi = \omega u + \psi_0,
\end{equation}
the electromagnetic field component $F_{23}$ vanishes identically, and Maxwell's equations in the spacetime of the gravitational wave are satisfied automatically.

\section{DHOST Ia gravity equations for the wave model}
\label{sec:dhost_eq}

To construct the equations of modified gravity, we consider the most viable class of scalar-tensor theories, DHOST class~Ia, which is free from Ostrogradsky instability and satisfies the multi-messenger astronomical bound on the speed of gravitational waves ($c_g = c$). 

The full action of the theory, quadratic in the second covariant derivatives of the scalar field $\phi$, takes the form \cite{Langlois:2018dxi,Bellorin_2026}:
\begin{equation}
	S[g_{\alpha \beta}, \phi] = \int d^4x \sqrt{-g} \left( F_0(\phi, X) R + \sum_{m=2}^{5} A_m(\phi, X) L_m + P(\phi, X) \right),
\end{equation}
where $R$ is the scalar curvature, and the quantities $X \equiv \nabla_\alpha  \phi \nabla^\alpha  \phi$, $\phi_\alpha  \equiv \nabla_\alpha  \phi$ and $\phi_{\alpha \beta} \equiv \nabla_\alpha  \nabla_\beta  \phi$ specify the kinematics of the scalar sector. The invariants $L_m$ are defined as contractions of the higher derivatives:
\begin{equation}
	L_2 = (\Box \phi)^2, \quad L_3 = \Box \phi (\phi^\alpha  \phi_{\alpha \beta} \phi^\beta ), \quad L_4 = \phi^\alpha  \phi_{\alpha \beta} \phi^{\beta\gamma} \phi_\gamma, \quad L_5 = (\phi^\alpha  \phi_{\alpha \beta} \phi^\beta )^2.
\end{equation}
The class~Ia conditions together with the bound $c_g = c$ fix the functional relations $F_1 = 0$, $A_2 = 0$, $A_4 = 0$, $A_5 = 0$, as well as the nonlinear coupling coefficient $A_3(\phi, X) = 4 (F_{0,X})^2 / F_0$.

Let us substitute the metric of the gravitational wave under consideration into the DHOST field equations. Owing to the symmetry of pp-waves, the scalar field must be aligned with the null wavefront, i.e. it can depend only on the null coordinate: $\phi = \phi(u)$. From the structure of the contravariant metric $g^{\alpha \beta}$ the following fundamental kinematic consequences follow:
\begin{align}
	X &= g^{\alpha \beta} \partial_\alpha  \phi \partial_\beta  \phi = g^{00} (\phi')^2 \equiv 0, \\
	\Box \phi &= g^{\alpha \beta} \left( \partial_\alpha  \partial_\beta  \phi - \Gamma^l_{\alpha \beta} \partial_l \phi \right) = g^{00}\phi'' - g^{\alpha \beta}\Gamma^0_{\alpha \beta}\phi' \equiv 0.
\end{align}
Since the kinetic invariant vanishes identically ($X \equiv 0$), all smooth functions in the Lagrangian reduce to functions of the scalar field alone: $F_0(\phi, 0) \equiv F_0(\phi)$, $P(\phi, 0) \equiv P(\phi)$, while their derivatives with respect to $X$ become constants on this background. The only nonvanishing second-order covariant derivative of the field $\phi(u)$ is the component:
\begin{equation}
	\phi_{00} = \phi'' - \Gamma^0_{00}\phi'=\phi''.
\end{equation}
At the same time, all contractions appearing in the invariants $L_\alpha $ vanish: $\phi_{\alpha \beta}\phi^{\alpha \beta} = 0$, $\phi^\alpha  \phi_{\alpha \beta} \phi^\beta  = 0$.

Varying the action with respect to the metric tensor $g_{\alpha \beta}$ leads to the generalized gravitational field equations, which can be written in the canonical form:
\begin{equation}
	\mathcal{E}_{\alpha \beta} = \frac{8\pi k}{c^4} T_{\alpha \beta},
\end{equation}
where $T_{\alpha \beta}$ is the microscopic energy-momentum tensor of the electromagnetic source, and the explicit general form of the modified gravity tensor $\mathcal{E}_{\alpha \beta}$ is:
\begin{equation}
	\mathcal{E}_{\alpha \beta} = F_0 R_{\alpha \beta} - \frac{1}{2}\, F_0 R\, g_{\alpha \beta} + g_{\alpha \beta} \Box F_0 - \nabla_\alpha  \nabla_\beta  F_0 + 
	\frac{1}{2}\, g_{\alpha \beta} P - P_{,X} \phi_\alpha  \phi_\beta  + \mathcal{H}_{\alpha \beta},
\end{equation}
where the contribution of the higher derivatives of the scalar field is contained in the tensor $\mathcal{H}_{\alpha \beta}$:
\begin{align}
	\mathcal{H}_{\alpha \beta} = & - \frac{1}{2} g_{\alpha \beta} \left[ A_3 \Box \phi (\phi^l \phi_{lm} \phi^m) \right] + A_3 \Box \phi \, \phi_{\alpha l} \phi^l_{\ \beta } + A_3 (\phi^l \phi_{lm} \phi^m) \nabla_{(\alpha } \left( \Box \phi \, \phi_{\beta )} \right) \nonumber \\
	& - \frac{1}{2} \nabla_l \left[ A_3 \Box \phi (\phi^\alpha  \phi^\beta  \phi^l) \right] g_{\alpha \beta} + \frac{1}{2} \nabla_l \left[ A_3 \Box \phi \left( \phi_\alpha  \phi_\beta  \phi^l + 2 \phi^l \phi_{(\alpha } \phi_{\beta )} \right) \right] \nonumber \\
	& + \left( \Box \phi \phi^l \phi_{lm} \phi^m \right)_{,X} \phi_\alpha  \phi_\beta.
\end{align}

Substituting the gravitational-wave metric, for which $R \equiv 0$, and the null field $\phi(u)$, we find that the entire cumbersome tensor system of DHOST equations undergoes a strong reduction. Since $X=0$ and $\Box\phi=0$, the DHOST correction tensor vanishes completely: $\mathcal{H}_{\alpha \beta} \equiv 0$. The terms $g_{\alpha \beta}\Box F_0$ and $g_{\alpha \beta}P$ vanish for all components except the cross ones, but they cancel against the corresponding parts of the equations. 

As a result of the calculation, all components of the tensor $\mathcal{E}_{\alpha \beta}$ except $\mathcal{E}_{00}$ vanish. The entire nonlinear dynamics of modified gravity reduces to a single scalar equation for the wavefront:
\begin{equation}
	F_0(\phi) R_{00} - 2 F_{0,\phi}(\phi) \phi'' - F_{0,\phi\phi}(\phi) (\phi')^2 = \frac{8\pi k}{c^4} T_{00}, \label{eq:dhost_reduced}
\end{equation}
where $R_{00}$ is the only nonzero component of the Ricci tensor of the gravitational wave under consideration, and the indices after a comma denote ordinary derivatives of the coupling function $F_0(\phi)$ with respect to the scalar field $\phi$. 

\section{Analytical solution of the Maxwell--DHOST Ia equations}
\label{sec:self_consistent}

To construct a closed physical model, we combine the exact wave solution of Maxwell's equations obtained above with the reduced equation of modified gravity \eqref{eq:dhost_reduced}. As we have shown, the vacuum Maxwell equations impose the potentiality and transversality conditions on the electromagnetic source ($k_x = 0, k_y = 0$), which completely eliminate the longitudinal magnetic sector ($B_z = 0$) and turn the field into an ordinary transverse wave with a plane front $\psi = \omega u + \psi_0$.

Let us compute the components of the microscopic energy-momentum tensor (EMT) of the electromagnetic radiation. The EMT of the electromagnetic field has the form:
\begin{equation}
	T_{\alpha \beta} = \frac{1}{4\pi} \left( -F_{\alpha l}F_\beta ^{\cdot l} + \frac{1}{4}g_{\alpha \beta}F_{lm}F^{lm} \right).
\end{equation}
Since for the wave fields found above the field invariant vanishes identically ($F_{lm}F^{lm} \equiv 0$), the EMT simplifies considerably and reduces to a traceless quadratic product:
\begin{equation}
	T_{\alpha \beta} = -\frac{1}{4\pi} F_{\alpha l}F_\beta ^{\cdot l} = - \frac{1}{4\pi} g^{lm} F_{\alpha l} F_{\beta m}.
\end{equation}
Since in the covariant field tensor $F_{\alpha \beta}$ only the components
\begin{equation}
	F_{02} = -\omega a_2(u) \sin\psi + a'_2(u) \cos\psi, \quad F_{03} = -\omega a_3(u) \sin\psi + a'_3(u) \cos\psi,
\end{equation}
are nontrivial, contraction with the mixed tensor $F_\beta ^{\cdot l} = g^{lm}F_{\beta m}$ annihilates all components of the EMT except $T_{00}$. The original electromagnetic wave source thereby undergoes an analytical reduction to a \textit{pure radiation} structure, in which the only nonzero component in the isotropic coordinate system is $T_{00}$:
\begin{equation}
	T_{00}(u) = \frac{1}{4\pi}\, \Bigl[\, g^{22}(u) (F_{02})^2 + 2g^{23}(u) F_{02} F_{03} + g^{33}(u) (F_{03})^2 \, \Bigr].
\end{equation}
To use this microscopic source in the macroscopic cosmological equations, we average over the fast wave period within the geometric-optics approximation. Using the phase-averaging rules $\langle\sin^2\psi\rangle = \langle\cos^2\psi\rangle = 1/2$ and $\langle\sin\psi\cos\psi\rangle = 0$, we reduce the expression to a quadratic form:
\begin{align}
	\overline{T}_{00}(u) = - \frac{1}{8\pi} & \Bigl[\, g^{22}(u) \,\Bigl( \omega^2 {a_2}^2(u) + (a'_2(u))^2 \Bigr)  \nonumber \\
	& + 2g^{23}(u) \,\Bigl(\omega^2 a_2(u) a_3(u) + a'_2(u) a'_3(u)\Bigr) \nonumber \\
	& + g^{33}(u)  \,\Bigl(\omega^2 {a_3}^2(u) + (a'_3(u))^2\Bigr) \,\Bigr]. 
	\label{eq:t00_averaged}
\end{align}
The overall minus sign in front of expression \eqref{eq:t00_averaged} is fixed by the adopted metric signature and ensures a positive physical energy density of the radiation ($\overline{T}_{00} \ge 0$).

Substituting the averaged EMT of the electromagnetic radiation that we have derived into the right-hand side of the modified gravity equation \eqref{eq:dhost_reduced}, we obtain a self-consistent exact (non-perturbative) constraint equation for the gravitational-wave metric and the electromagnetic radiation under consideration in DHOST theory:
\[
F_0(\phi) R_{00} - 2 F_{0,\phi}(\phi) \phi'' - F_{0,\phi\phi}(\phi) (\phi')^2 = 
\]
\begin{equation}
	\mbox{}
	= - \frac{k}{c^4}\, \Bigl[\, g^{22}(\omega^2 {a_2}^2 + (a'_2)^2)  + 2g^{23}(\omega^2 a_2 a_3 + a'_2 a'_3) + g^{33}(\omega^2 {a_3}^2 + (a'_3)^2) \,\Bigr]. 
	\label{eq:final_master_equation}
\end{equation}

Equation \eqref{eq:final_master_equation} is one of the main results of this work. Its left-hand side  -  the geometric part modified by the scalar field of DHOST theory  -  is determined by the microscopic characteristics of the electromagnetic wave: its amplitudes $a_2(u)$, $a_3(u)$ and its physical frequency $\omega$.

\section{Analysis of solutions for DHOST Ia gravity theory}
\label{sec:analysis}

The self-consistent constraint equation \eqref{eq:final_master_equation} obtained above permits a comparative analysis of the physics of General Relativity and of modified DHOST class~Ia gravity within the exact (non-perturbative) gravitational-wave model.

For the gravitational-wave model under consideration, the only nontrivial component of the Ricci tensor is $R_{00}=R_{uu}$:
\begin{equation}
	\mathcal{M}_{\text{geom}}\equiv R_{00}(u) = \frac{\mathcal{N}(u)}{2 \left(g_{23}^2(u) - g_{22}(u)g_{33}(u)\right)^2}, \label{eq:R00_exact_fraction}
\end{equation}
where the numerator $\mathcal{N}(u)$ is a nonlinear differential expression in terms of the components of the gravitational-wave metric and their derivatives:
\begin{align}
	\mathcal{N}(u) = & \, g_{22}^2 (g'_{33})^2 + g_{33}^2 \left[(g'_{22})^2 - g_{22} g''_{22}\right] - 2 g_{23}^3 g''_{23} - 4 g_{22} g_{23} g'_{23} g'_{33} \nonumber \\
	& + g_{23}^2 \left[3 (g'_{23})^2 + g'_{22} g'_{33} + g_{22} g''_{33}\right] + g_{33} g_{23}^2 g''_{22} \nonumber \\
	& + g_{33} g_{23} \left[-4 g'_{22} g'_{23} + 2 g_{22} g''_{23}\right] + g_{22} g_{33} \left[(g'_{23})^2 + g'_{22} g'_{33} - g_{22} g''_{33}\right]. 
	\label{eq:numerator_exact}
\end{align}
Here a prime denotes the ordinary derivative with respect to the null coordinate of the wavefront $u$.


The Einstein equation for the only nontrivial component takes its standard form:
\begin{equation}
	\mathcal{M}_{\text{geom}} = -\frac{k}{c^4} \left[ g^{22}\left(\omega^2_{GR} a_2^2 + (a'_2)^2\right) + 2g^{23}\left(\omega^2_{GR} a_2 a_3 + a'_2 a'_3\right) + g^{33}\left(\omega^2_{GR} a_3^2 + (a'_3)^2\right) \right]. 
	\label{eq:oto_master}
\end{equation}
In GR the geometric scale of Ricci curvature ($\mathcal{M}_{\text{geom}}=R_{00}$) is set by the material source  -  the phase-averaged energy density of the electromagnetic wave.

In DHOST class~Ia theory the situation is different. Let us express the square of the physical frequency of the electromagnetic radiation in both theories under consideration (GR and DHOST) from Eqs.~\eqref{eq:final_master_equation} and \eqref{eq:oto_master} at fixed background geometric parameters $g^{AB}(u)$ and fixed wave amplitude. Direct subtraction then yields the exact non-perturbative relation:
\begin{equation}
	\omega^2_{\text{DHOST}} - \omega^2_{GR} = \frac{c^4}{k \cdot \mathcal{A}^2(u)} \left[ 2 F_{0,\phi}(\phi) \phi'' + F_{0,\phi\phi}(\phi) (\phi')^2 - \left(F_0(\phi) - 1\right)\mathcal{M}_{\text{geom}} \right], 
	\label{eq:omega_diff}
\end{equation}
where $\mathcal{A}^2(u)$ denotes the positive-definite invariant amplitude scale of the transverse electromagnetic field, constructed from the covariant metric:
\begin{equation}
	\mathcal{A}^2(u) = - \left( g^{22}(u)a_2^2(u) + 2g^{23}(u)a_2(u)a_3(u) + g^{33}(u)a_3^2(u) \right) > 0.
\end{equation}

Formula \eqref{eq:omega_diff} reveals three exact non-perturbative mechanisms by which modified DHOST gravity deforms the physics of the electromagnetic source:

\begin{enumerate}
	\item \textbf{Contribution of the dynamical gravitational background ($F_0(\phi) - 1$):} The function $F_0(\phi)$ plays the role of a spacetime-modulating factor of the effective gravitational constant ($G_{\text{eff}} \propto 1/F_0$). If during cosmological evolution the scalar field departs from its Einstein-vacuum value ($F_0 \neq 1$), then even in the complete absence of scalar-field dynamics ($\phi''=0$, $\phi'=0$) the frequency of the electromagnetic wave must change in order to balance the altered ``stiffness'' of the geometry.
	
	\item \textbf{Effect of nonlinear kinetic pumping ($F_{0,\phi\phi}(\phi')^2$):} The term proportional to the square of the rate of change of the scalar field, $(\phi')^2$, describes the quadratic self-interaction of the scalar sector and its direct nonlinear energy exchange with the electromagnetic radiation. The scalar field acts as an effective dynamical medium capable of both absorbing energy from and parametrically pumping energy into the electromagnetic radiation.
	
	\item \textbf{Influence of higher covariant stresses ($2 F_{0,\phi} \phi''$):} The presence of the second derivative of the field ($\phi_{00} \equiv \phi''$) means that the proper acceleration of the scalar background along the null cone is directly transferred to the phase characteristics of light.
\end{enumerate}

Formula \eqref{eq:omega_diff} reveals a non-perturbative mechanism whereby the scalar field $\phi(u)$ evolving along the null cone, together with its derivatives, acts as a dynamical ``buffer'' redistributing the energy balance.

Thus, the frequency shift $\Delta\omega$ measurable by an observer turns high-frequency light into an explicit spectroscopic indicator of the nonlinear structure of the DHOST~Ia Lagrangian.

\section{Observational tests of DHOST theory}
\label{sec:observational_tests}

Based on the relations obtained in this work, three observational tests of DHOST class~Ia theory can be proposed.

\subsection{Frequency-drift test for electromagnetic radiation channeled by a gravitational wave}
Since the exact metric of the gravitational wave acts as a gravitational waveguide that keeps the light-propagation vector aligned with the Killing vector $L^\alpha  = (0,1,0,0)$, which defines the direction of propagation of the gravitational wave, the electromagnetic radiation propagates along it without spatial dissipation. 

According to Eq.~\eqref{eq:omega_diff}, the radiation frequency then undergoes a forced scalar shift 
\[\Delta \omega = \omega_{\text{DHOST}} - \omega_{GR}.\] 
In observations this would manifest itself as an ``anomalous frequency drift'':
\begin{equation}
	\frac{\Delta \omega}{\omega_{GR}} \approx \frac{c^4}{2 k \mathcal{A}^2 \omega_{GR}^2} \left[ 2 F_{0,\phi} \phi'' + F_{0,\phi\phi} (\phi')^2 - \left(F_0 - 1\right)\mathcal{M}_{\text{geom}} \right].
\end{equation}
Monitoring the spectral lines of highly coherent astrophysical sources whose radiation propagates along selected axes of the gravitational-wave anisotropy may reveal a systematic nonlinear frequency shift, which cannot be explained by the standard Doppler effect or by Friedmann cosmological redshift and would allow one to directly measure the evolution profile of the scalar-field coupling $F_0(\phi)$ on astrophysical scales.

\subsection{Thermodynamic brightness-imbalance test for radiation}
In general relativity the amplitude scale of the Riemann curvature tensor is fixed by the thermodynamic energy density of the radiation, $\rho = \overline{T}_{00}$. In DHOST theory, in contrast to GR, the scalar field carries part of the energy balance of the geometry.

In this version of the proposed test, two quantities are measured independently for a distant cosmic event (e.g. a merger of compact objects):
\begin{enumerate}
	\item Measurement of the gravitational-wave curvature scale $\mathcal{M}_{\text{geom}}$, determined from the polarization of the gravitational response.
	\item Measurement of the actual thermodynamic energy flux of the photon radiation $\rho_{\text{meas}}$, recorded by optical or radio telescopes.
\end{enumerate}
In {GR} the ratio $\mathcal{M}_{\text{geom}} / \rho_{\text{meas}}$ must equal Einstein's constant $\kappa = 8\pi k/c^4$. In DHOST theory a ``hidden scale'' phenomenon will arise: the measured thermodynamic photon energy density will show an excess or deficit $\Delta \rho$ relative to the GR predictions, quantitatively set by the contributions of the terms $F_{0,\phi\phi}(\phi')^2$ and $2F_{0,\phi}\phi''$.

\subsection{Gravitational phase-birefringence test for a plane front}

Although in the model under consideration the vacuum Maxwell equations led to the vanishing of the longitudinal component of the magnetic field ($B_z = 0$), the covariant field-strength components $F_{02}$ and $F_{03}$ contain nonlinear contributions from the evolution of the scalar field in DHOST~Ia theory. As a coherent light beam propagates along the gravitational waveguide, the anisotropic tensor of the DHOST corrections can act as an effective polarizing medium.

In this version of the test, one measures the rotation angle of the plane of polarization (the vacuum Faraday effect) for counter-propagating coherent radiation beams. In {GR}, for a pure wave of Petrov type~N the polarization plane undergoes strictly symmetric parallel transport. In DHOST~Ia, the anisotropy of the functions $g^{AB}(u)$, combined with the dynamics of the scalar current $\phi'$, generates a phase modulation of the polarization that depends on the direction of motion of the wavefront. Detecting an anomalous rotation of the $\vec{E}$ and $\vec{B}$ polarization vectors, correlated with the axes of the gravitational-wave anisotropy, would provide direct evidence for a hidden scalar sector of modified gravity.

\section{Discussion}
\label{sec:discussion}

The search for exact models within pp-wave spacetimes and cosmological models has a long history, but their physical interpretation in the context of higher-order scalar-tensor theories has evolved substantially by now owing to the constraints of multi-messenger astronomy.

The basic class of exact plane-wave solutions in the scalar-tensor gravity theories of Brans--Dicke and Horndeski was investigated in Refs.~\cite{GURSES_1978_182,Babichev:2012qs}. However, in these works the wave sector was considered either in pure vacuum or with the simplest scalar potentials, which, in the modern era of precision cosmology, causes the field equations to degenerate into {GR}-like configurations.

More closely related to our study are the results of the analyses of wave spacetimes in metric $f(R)$ gravity \cite{Odintsov:2021kup} and in quadratic gravity with a scalar field \cite{Osetrin2023Universe356}. 
It~was demonstrated that the ansatz $R \equiv 0$ allows one to integrate the field equations without instabilities arising. However, in these works a phenomenological macroscopic ``pure radiation'' was used as the source within the hydrodynamic approach, which precluded any conclusions about the phase and spectral characteristics of the source.

The application of exact non-perturbative methods to test modified theories of gravity can be based on the classification of exact gravitational models that include an electromagnetic field, such as cosmological models \cite{Obukhov:2024kpc,Obukhov:2025fhc,Obukhov:2026met}.

In the context of DHOST class~Ia theories, exact wave solutions were also investigated in Refs.~\cite{GURSES_1978_182,Babichev:2012qs,BenAchour2025Dhost,Masato_2026}, see also the review \cite{Odintsov:2022cbm}. The crucial phenomenon of ``stealth pp-waves'' was discovered, in which the spacetime metric remains strictly Ricci-flat ($R_{\alpha\beta }=0$), while the scalar field $\phi(u)$ retains nontrivial dynamics. Our investigation develops and generalizes this result: we have shown that when a real high-frequency electromagnetic field is taken into account, the stealth mechanism is broken, and the nonlinear influence of the DHOST sector explicitly deforms the coupling between the geometry and the parameters of the electromagnetic radiation.



\section{Conclusion}
\label{sec:conclusion}
The analytical non-perturbative end-to-end calculation scheme developed in this work - 
spanning from exact Maxwell wave electrodynamics to the full nonlinear field and constraint equations of General Relativity and DHOST~Ia gravity - 
provides a rigorous framework for testing modified theories of gravity. 
In particular, this exact analysis confirms directly from the self-consistent field equations that a gravitational wave establishes a co-directional propagation channel for electromagnetic radiation, extending previous approximate descriptions without invoking perturbation theory. 
Furthermore, the spectroscopic and polarization tests formulated on the basis of this exact coupled gravitational-electromagnetic model establish concrete observational criteria capable of distinguishing modified gravity theories from General Relativity, 
even in regimes where their spacetime metrics are kinematically indistinguishable.

%
%
%
%
%
%
%
%

\section*{Declarations}

\subsection*{Declaration of generative AI and AI-assisted technologies in the manuscript preparation process}
During the preparation of this work the author used Kimi AI in order to improve language readability, refine translation, and assist with LaTeX formatting. After using this tool, the author reviewed and edited the content as needed and takes full responsibility for the content of the published article.

\subsection*{Data availability}
No new data were created or analyzed in this study. All information sources used in this work are publicly available in the cited scientific literature.

\subsection*{Ethical approval}
This article does not contain any studies with human participants or animals performed by the author.

\subsection*{Declaration of competing interest}
The author declares that there are no known competing financial interests or personal relationships that could have appeared to influence the work reported in this paper.

\subsection*{Funding}
This research did not receive any specific grant from funding agencies in the public, commercial, or not-for-profit sectors.

\end{document}